\documentclass[final]{ustcstep}

\usepackage{color}
\usepackage{abstract}
\usepackage{times}
\usepackage{multicol}
\usepackage{graphics}
\usepackage[figuresleft]{rotating}
\usepackage{lscape}
\usepackage{bbding}
\usepackage{amssymb}
\usepackage{array}
\usepackage{textcmds}
\usepackage{float}
\usepackage{ulem}
\usepackage[square]{natbib}
\usepackage[]{times, graphicx}
\def\gsim{\;\lower4pt\hbox{${\buildrel\displaystyle >\over\sim}$}\;}
\def\lsim{\;\lower4pt\hbox{${\buildrel\displaystyle <\over\sim}$}\;}
\def\grls{\;\lower4pt\hbox{${\buildrel\displaystyle >\over <}$}\;}

\newcommand\addr[2]{{\footnotesize \it $^{#1}$#2}\\}
\def\grad{$^\circ$}
\def\kms{km s$^{-1}$}
\def\acc{m s$^{-2}$}
\def\th{$^{th}$}

\def\apj{ApJ}
\def\apjl{ApJL}

\def\solphys{Solar Phys.}
\def\aap{A\&A}

\def\pasj{PASJ}
\def\ssr{SSR}

\begin{document}

\title{On the Observation and Simulation of Solar Coronal Twin Jets}

\author{Jiajia Liu$^{1,3,*}$, Fang Fang$^2$, Yuming Wang$^{1,5}$, 
Scott W. McIntosh$^{4}$, Yuhong Fan$^{4}$, Quanhao Zhang$^{1,3}$ \\
\addr{1}{CAS Key Laboratory of Geospace Environment, School of Earth and Space Sciences, 
University of Science and Technology of China, Hefei, Anhui 230026, China}
\addr{2}{Laboratory For Atmospheric and Space Physics, University of Colorado at Boulder, 
1234 Innovation Dr., Boulder, CO 80303, USA}
\addr{3}{Collaborative Innovation Center of Astronautical Science and Technology, Hefei 230026, China}
\addr{4}{High Altitude Observatory, National Center for Atmospheric Research, 
P.O. Box 3000, Boulder, CO 80307, USA.}
\addr{5}{Synergetic Innovation Center of Quantum Information \& Quantum Physics, 
University of Science and Technology of China, Hefei, Anhui 230026, China}
\addr{*}{Corresponding Author, Contact: ljj128@ustc.edu.cn}}

\maketitle
\tableofcontents

\begin{abstract}

We present the first observation, analysis and modeling of solar coronal {\it twin jets}, which occurred 
after a preceding jet. Detailed analysis on the kinetics of the preceding jet reveals its blowout-jet 
nature, which resembles the one studied in \emph{Liu et al. 2014}. However the erupting process and kinetics 
of the twin jets appear to be different from the preceding one. In lack of the detailed information on the magnetic fields in 
the twin jet region, we instead use a numerical simulation using a three-dimensional (3D) MHD model as described in \emph{Fang et al. 2014}, 
and find that in the simulation a pair of twin jets form due to reconnection between the ambient open fields and 
a highly twisted sigmoidal magnetic flux which is the outcome of the further evolution of the magnetic fields 
following the preceding blowout jet. Based on the similarity between the synthesized and observed emission we 
propose this mechanism as a possible explanation for the observed twin jets. Combining our observation and simulation, 
we suggest that with continuous energy transport from the subsurface convection zone into the corona, solar coronal 
twin jets could be generated in the same fashion addressed above.

\end{abstract}

\section{Introduction}\label{sect:intr}

Decades have passed since the first observations on solar jets 
\citep[named as surges in ][]{Newton1934}, which are thought to 
play an important role in solar wind acceleration and coronal heating 
\citep[e.g.,][]{Tsiropoula_Tziotziou2004, Tian2014}. A generalized definition 
of solar jets includes the terms of H$\alpha$ surges 
\citep[e.g.,][]{Canfield1996, Jibben_Canfield2004}, UV/EUV/X-ray jets 
\citep[e.g.,][]{Schmieder1988, Patsourakos2008, Tian2014, Liu2015} and spicules 
\citep[e.g.,][]{DePontieu2007, Shibata2007}, among which their different 
names come from different dominant temperatures and sizes. As shown 
in many previous works \citep[][as a review]{Shibata1996}, different jets 
obtain quite different physical characteristics such as length and axial 
speed, which range from few to hundreds megameters and ten to thousands 
kilometers per second, respectively.

Despite the different properties of different jets, it is believed that 
they are triggered by the similar mechanism 
\citep[except type I spicules,][]{DePontieu2007}. Reconnections between 
newly emerging twisted loops with pre-existing ambient open fields \citep[e.g.,][]{Moreno-Insertis2008} 
lead to the heating and initiation of bulks of plasma, which are observed as materials of a jet 
\citep{Savcheva2007}. Twists transferred from the 
emerging flux then lead to the rotational motion of jets, as observed and studied 
widely in observation and simulation \citep[e.g.,][]{Xu1984, Shibata_Uchida1985, 
Canfield1996, Shimojo2007, Pariat2010, Liu2012, Liu2014, Fang2014}.

Although the triggering mechanism has been studied comprehensively and thoroughly, 
further evolution of the system after the reconnection and the detailed energy budget 
during jet events still stays unclear. As known by the community, during a jet event, 
the magnetic free energy is released through two ways. One is reconnection and the 
other is post-reconnection relaxation of the magnetic field structure, which is always 
manifested by the rotational motion of a jet. Observation employing the unprecedented 
combination of the {\it {\it SDO}} \citep{Pesnell2012} and {\it STEREO} \citep{Kaiser2008} data of 
a solar EUV jet by \cite{Liu2014} has demonstrated that the continuous relaxation 
of the post-reconnection magnetic field structure is an important energy source for a jet 
to climb up higher than it could through only reconnection. Analysis in \cite{Liu2014} shows that the kinetic energy of 
the jet gained through the relaxation is about 1.6 times of that gained from the 
reconnection. The importance of the post-reconnection relaxation process which introduces 
upward Lorentz force has also been demonstrated in the 3D MHD numerical simulation work 
by \cite{Fang2014}.

Further releasing of magnetic free energy may take place in terms of recurring jets. When persistent flow 
continuously injects energy into the corona from the sub-surface regions, recurring jets may present 
\citep{Pariat2010}. However, in this paper, we will present the observation and analysis on another 
possibility - solar coronal twin jets after a preceding blowout jet. After the detailed analysis on the kinetics, 
we continue the 3D MHD simulation work in \cite{Fang2014} where a sub-photospheric buoyant magnetic 
flux rope emerges into the corona and reconnects with the ambient fields, producing a blowout jet, and 
demonstrate that the twin jets are generated via the reconnection between the ambient open fields 
with a highly twisted sigmoidal magnetic flux which is the outcome of the evolution of the magnetic 
field configuration during the preceding jet. Based on the observations in Section 2 and 3, and the 
simulation in Section 4, we present our conclusions and discussions in the last section

\section{Overview on Observations}\label{sect:obs}

\begin{figure}[tbh!]
\includegraphics[width=0.9\hsize]{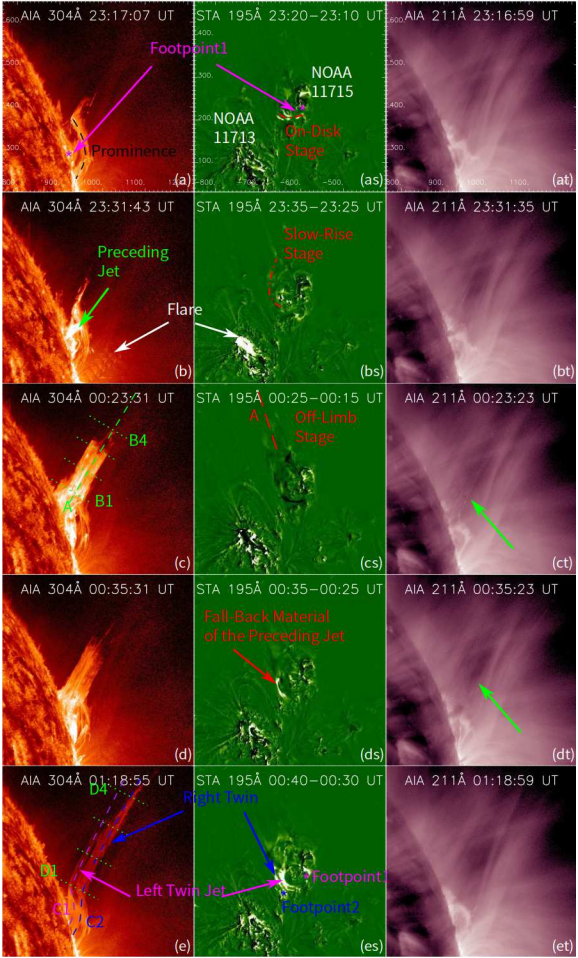}
\caption{Time-sequential observations on the analyzed event. (a) and (as): simultaneous 
observation on the on-disk stage of the preceding jet by {\it SDO} and {\it STEREO}. The purple 
asterisk shows the footpoint region (``Footpoint1") of the preceding jet.
The prominence is found not to interact with the jets and kept isolated from them. 
(b) and (bs): observation on the slow-rise stage of the preceding jet. (c) and (cs): observation 
when the preceding jet reached its maximum height. Slice ``A" is placed along the jet's 
axis to help analyze its axial motion, with slices ``B1" to ``B4" perpendicular 
to the axis. (d) and (ds): one moment during the descending phase of the preceding jet. 
(e) and (es): observation of {\it SDO} when the twin jets reached their maximum heights 
and of {\it STEREO} shortly after the twins were triggered. Slices ``C1" and ``C2" are placed 
along the jets' axes, with slices ``D1" to ``D4" perpendicular to the jets' axes. 
(at)-(et): simultaneous AIA 211 \AA \  observations with the green arrows 
indicating the hot component of the preceding jet.}\label{OV}
\end{figure}

The event focused on in this paper started around the midnight between April 10\th{} and the 
following day in the year 2013, performed an attractive show and ended after its 3-hour performance on the 
north-west limb of the Sun in the field-of-view (FOV) of the {\it Solar Dynamics Observatory ({\it {\it SDO}})}/AIA. A solar coronal jet 
surfed high into the solar corona before falling back to the solar surface. And another two jets 
came out around the end of the off-limb stage of it. The first jet could be found in all seven 
UV/EUV channels of {\it SDO}/AIA and in EUV observations by the {\it Solar TErrestrial RElations Observatory} ({\it STEREO})/EUVI, indicating the 
multi-thermal nature with a wide temperature range at least from ten thousands to ten 
millions in units of Kelvin (See online animation M1) \citep{Lemen2012}. However, as we can see from 
the 211 \AA \  observations in Figure~\ref{OV}, there were only few jet materials resolved 
and the jet was dominated by warm materials at the temperature of 304 \AA \  passband. Running-difference
images in animation M1 reveal that materials of the jets in different temperatures acted in the same way. 
A prominence could also be found at almost the same position seen from the {\it SDO} during the analyzed event 
(Figure~\ref{OV}(a)). However, as it is found to remain its main structure through out the whole event and 
not show any obvious interaction with the jets, we will not take it into account during our analysis 
in this paper.

The first jet, which will be referred to as the ``preceding jet" hereafter, originated from a 
footpoint region, says ``Footpoint1", located around E38\grad{}N18\grad{} within the active region 
NOAA 11715 in the FOV of {\it STEREO}-A (Figure\ref{OV}(as)). This location corresponded to W95\grad{} 
from the point of view of {\it SDO} taking the separation between these two probers of 133\grad{} into 
account, which means that some early (on-disk) evolution of the jet would hide from {\it SDO} 
observations - consistent with our temporal investigations on animation M1. The preceding jet 
was triggered around 23:17 UT April 10\th{}, and continuous chromospheric activities could be found around 
its footpoint region within the same active region before its excitation via 
{\it STEREO} observations. The jet traveled on disk for about 42 Mm until 
23:20 UT (red dashed curve in Figure~\ref{OV}(as)). It could only be seen as a small bright 
region during this on-disk stage from the {\it SDO}, shown as a purple asterisk in Figure~\ref{OV}(a) (and animation M2). 
Then, the jet was found to ascend slowly (slow-rise stage) to a height of 26 Mm in {\it SDO} 
observation (Figure~\ref{OV}(b)) and simultaneously continued its on-disk motion traveling 
for about 137 Mm in {\it STEREO} observation (shown as the red dashed curve Figure~\ref{OV}(bs)) before 
23:32 UT. Additionally, a C3.9 flare could also be observed in the neighboring active region 
NOAA 11713 (Figure~\ref{OV}(as)), which started at around 23:31 UT (Figure~\ref{OV}(b) and (bs)) 
during the jet's slow-rise stage. However, we'll not talk much into the relation between the 
jet and the flare as they were not in the same active region and their relations might be 
complicated which is beyond the topic of this paper.

After this slow-rise stage, the preceding jet began to ascend almost straightly off the 
limb with an average width of about 38 Mm (off-limb stage, Figure~\ref{OV}(c) and (cs)). 
The jet reached its maximum height at around 00:23 UT April 11\th{} (Figure~\ref{OV}(c)) 
and started to fall back to the solar surface after that. The falling-back materials 
could also be found in {\it STEREO} observations (Figure~\ref{OV} (ds)). Similar as the 
one studied in \cite{Liu2014}, rotational motion of the jet 
materials could be found over the whole off-limb stage of the jet.

Around the end of the off-limb stage of the preceding jet (00:35-00:40 UT), another two jets 
were triggered. They stayed so close in the FOV of {\it SDO} that they could be easily mistaken for one jet. However 
as we can see from {\it STEREO} observations, 
the left one originated from a footpoint region that was apparently around that of 
the preceding one, which located inside the active region NOAA 11715 (``Footpoint1", Figure~\ref{OV}(es)). 
While the right one was found to be originated from another footpoint region (``Footpoint2", Figure~\ref{OV} (es)), 
which was south of ``Footpoint1" and between the two active regions NOAA 11715 and 11713. Meanwhile, there was 
a gap between them with the distance of their axes about 20 Mm and widths of their tunnels about 14 Mm and 18 Mm respectively 
in {\it SDO} observations. Moreover, they finally reached different heights.

On the other hand, they are considered as ``twin jets" rather than ``two jets" due to: (1) they were simultaneously triggered 
with close footpoints and (2) they stayed very close together and shared plenty of 
commonalities in temperature, brightness and kinetic properties, as we will see in the rest of this paper. 
No obvious flux emergence could be found around the twins' footpoint regions through the observations of 
{\it STEREO} during their initiation. Meanwhile, only transient rotational motion of the twins could be found 
during their early off-limb stage. The rotational motion appeared to be much weaker than that of 
the preceding jet, with longer periods and much shorter lifetime. No rotational motion could be found thereafter.

\section{Kinetics}\label{sect:kinet}

One major concern when analyzing the kinetics of the observed jets is the projection effect, which may have pretty large 
influence on the resolved axial velocity of the materials within the jets. To correct the off-limb projection effect, 
we combine the observations from the two satellites ({\it SDO} and {\it STEREO}-A) employing a simple trigonometric analysis as 
done in \cite{Liu2014}. Knowing the separation of {\it SDO} and {\it STEREO}-A of about 133\grad{}, the projected 
length of the preceding jet at 00:23 UT of about 145 Mm in the view of {\it SDO} (Figure~\ref{OV}(c)) 
and 98 Mm at almost the same time in the view of {\it STEREO} (Figure~\ref{OV}(cs)), we find that the 
axis of the preceding jet was about 62\grad{} inward  
the plane of the sky in the view of {\it SDO}. The actual (physical) 
off-limb length and the axial speeds of the jet should be corrected with a factor 
of $\cos ^{-1} (62$\grad$) \approx 2.11$.

\subsection{Kinetics of the Preceding jet}\label{sect:prec}

\begin{figure}[tbh!]
\begin{center}
\includegraphics[width=0.9\hsize]{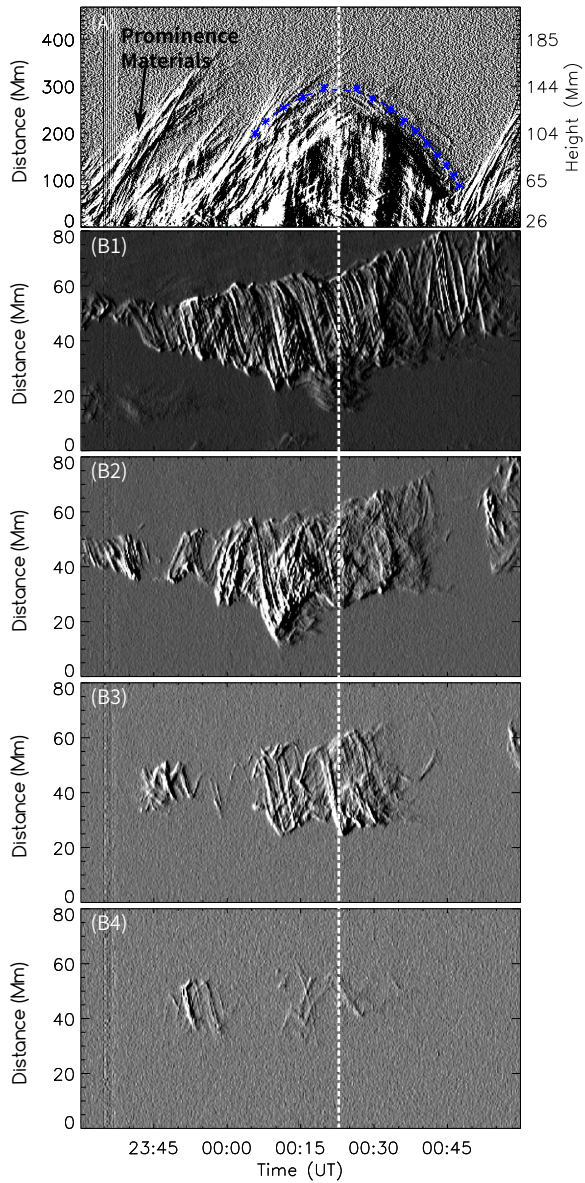}
\caption{(A): running-difference time-distance plot of slice ``A" in Fig~\ref{OV}(c) 
with the left y-axis distance along the slice after correcting the projection effect and the right y-axis height relative 
to the solar limb. The blue stars show one of the most 
distinguished tracks of the resolved sub-jets within the preceding jet, with the blue 
curve its parabolic fitting result. The black arrow indicates materials ejected from the prominence labeled in Fig~\ref{OV}(a).
(B1)-(B4): running-difference time-distance plots of slice ``B1" to ``B4" in Fig~\ref{OV}(c). 
Continuous rotational motion could be found in all the four panels with roughly 
the same period. The vertical dashed white line shows the time when the jet reached 
its maximum height.}\label{DPRE}
\end{center}
\end{figure}

To investigate the kinetics of the preceding jet during its off-limb stage, 
we place a slice ``A" along its axis as shown in Figure~\ref{OV}(c) with a width 38 Mm 
(same as that of the jet). The running-difference time-distance diagram derived from 
slice ``A" is shown in Figure~\ref{DPRE}(A).

Sub-jets expelled successively as parts of the preceding jet could be found as 
bright-dark alternating stripes in Figure~\ref{DPRE}(A). As indicated by the 
black arrow, materials before the preceding jet were not sub-jets of the preceding jet - they were 
ejactas from the prominence labeled by the black 
dashed line in Figure~\ref{OV}(a). Tracing the trajectories of 
several sub-jets in the time-distance plot to investigate the very detailed kinetics 
of them \citep[as done in][]{Liu2014} turns out to be impossible for this particular 
event, due to the poor contraries between these sub-jets and their intermissions. 

These sub-jets turned out to have an upward acceleration below a certain height 
and then started to decelerate. The exact value of the acceleration of the sub-jets near the bottom 
can not be well estimated, due to the bad contraries and that different 
sub-jets seemed to yield different values. However, the downward acceleration of these sub-jets 
during the late ascending phase turned out to be similar with that during the descending phase. 
Blue stars in Figure~\ref{DPRE}(A) shows the trajectory of one of the sub-jets, which can be well fitted 
by a parabolic function, shown as the blue dashed curve in Figure~\ref{DPRE}(A). The downward acceleration 
obtained by the parabolic fit is about -181.8$\pm$9.6 \acc{}. We can conclude that similar as 
the one studied in \cite{Liu2014}, sub-jets of the preceding jet experienced an 
``accelerating-decelerating-falling" process during its off-limb stage. And the downward 
acceleration during the late ascending phase and whole descending phase are both less then 
the average gravity (-220 \acc{}).

Continuous rotational motion could also be observed during the whole off-limb 
stage of the preceding jet. Four parallel slices named ``B1" to ``B4" are placed 
at four different distances (36 Mm, 126 Mm, 216 Mm and 306 Mm) perpendicular to the 
jet's axis to help probe the rotational motion (Fig~\ref{OV}(c)). All the four slices 
are oriented toward lower latitudes. The resulting running-difference time-distance 
plots of them are shown in Figure~\ref{DPRE}(B1) to (B4). Besides the expanding 
(Figure~\ref{DPRE}(B1)) and swinging-toward-lower-latitude (Figure~\ref{DPRE}(B1)-(B3)) 
motion of the jet, bundles of threads rotating around the jet's axis could also 
be observed easily via the sine-like tracks in the time-distance plots. Although there are 
not strong signal of opposite propagating motion for some threads in Figure~\ref{DPRE}(B1) due to shadowing effect 
of optically thick materials, turning motion of them around the edges, continuous investigation from (B2) 
to (B4) and animation M2 together can indicate us their rotational motion.
As seen from the diagrams, the rotational motion kept almost the same period at these four 
different heights and didn't show obvious deceleration after the jet reaching the 
maximum height (the white dashed line indicates the time when the jet reached 
the maximum height). 

Analysis employing a sine-function fit along these sine-like tracks on 
the four time-distance plots reveals the periods of the rotational motions at 
these four different heights of $564\pm79\ s$, $586\pm46\ s$, $557\pm147\ s$ 
and $606\pm35\ s$ respectively. These similar values again support our visual 
observation that the rotational motion kept almost the same behavior at different 
heights and didn't apparently slow down with time. As the last slice gives 
only few tracks that may lead to large errors, we average the periods obtained 
from the first three slices to estimate the average period of the rotational 
motion, which turns out to be 569 s (with an error 91 s). Taking the average 
width 38 Mm of the jet into account, the average line speed of the rotational 
motion during the off-limb stage is then estimated to be about $209.8\pm33.6$ \kms{}. 
As the upward Lorentz force acting on the jet after reconnection was resulted from 
the untwisting motion \citep{Shibata_Uchida1985}, the nearly invariable rotating period then is 
consistent with the above result that the downward acceleration in the late ascending phase and 
in the descending phase were almost the same.

\subsection{Kinetics of the Twin jets}\label{sect:twin}

\begin{figure}[tbh!]
\begin{center}
\includegraphics[width=0.9\hsize]{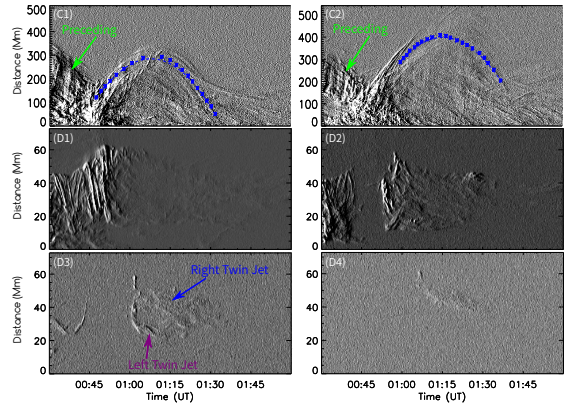}
\caption{(C1) and (C2): running-difference time-distance plots for slice ``C1"
and ``C2" in Fig.~\ref{OV}(e) along the axis of the left and right twin jet 
respectively. The blue stars represent trajectories 
of materials within the twins, with the blue curves their parabolic fitting 
results. (D1) to (D4): running-difference time-distance plots for 
slices ``D1" to ``D4" perpendicular to the twins' axes as shown in Fig.~\ref{OV}(e). Only very weak rotational 
motion of the twins could be found at their early off-limb stage from these four time-distance plots.}\label{TWIN}
\end{center}
\end{figure}

Similar method reveals almost the same inclination of the twins' axes 
as the preceding jet related to the plane of the sky.
To probe the axial motions of the twin jets, we place two slices along their 
axes labeled as ``C1" and ``C2" in Figure~\ref{OV}(e) for the left and right 
twin jet respectively. Their corresponding running-difference time-distance plots 
are shown in Figure~\ref{TWIN}(C1) and (C2) with the y-axis the distance along 
the slices after correcting the projection effects.

As in Figure~\ref{TWIN}(C1), the left twin jet reached a maximum 
distance along the slice of about 330 Mm at around 01:09 UT. The downward 
acceleration during its ascending phase and descending phase turned out 
to be similar. Parabolic fitting (blue dashed curve in Fig.~\ref{TWIN}(C1)) on one of the 
trajectories (blue stars in Fig.~\ref{TWIN}(C1)) gives us an average deceleration of about 
-228.9$\pm$8.1 \acc{}. It is hard to determine the maximum distance of 
the right twin jet (Fig.~\ref{TWIN}(C2)) - it seemed to reach a distance that was 
beyond the FOV of {\it SDO} observations or faint to be invisible. However the dynamics 
seems to be similar as the left one with the average downward acceleration of about 
-234.5$\pm$10.7 \acc{} (one of the trajectories is shown as the blue stars with 
the parabolic fitting result the blue dashed curve in Fig.~\ref{TWIN}(C2)). All of the above values 
resemble the average local gravity (-220 \acc{}), which indicates that there was 
almost no extra driving force acting on the twins.

Meanwhile, we place four parallel slices ``D1" to ``D4" at different heights 
perpendicular to the twins' axes to investigate the rotational motion of 
the twins if there was any. Corresponding running-difference time-distance 
plots are shown in Figure~\ref{TWIN}(D1) to (D4) with the y-axis the distance 
along the slices from higher latitudes. It is hard to find the twins in 
panel (D1) because the preceding jet was much brighter than the twins. 
Very weak rotational motions with very few turns, long period, low 
velocity and small amplitude of the twins during their early ascending 
phase could be found in panel (D2). However, it is hard to 
distinguish them from each other in this panel because they were too close 
together. The twins became more distinct 
from each other in panel (D3) and the left twin jet disappeared in panel 
D4 because slice D4 is located at a height beyond the left twin jet could 
reach. Then, we can conclude from these four panels that the twins 
only showed very weak rotational motion during their early off-limb stages,
which is very different from the preceding one.

\section{Simulation}\label{sect:simu}

\cite{Fang2014} studied the eruption of coronal jets by numerical simulations on 
the interaction between emerging and pre-existing magnetic fields. A stationary 
central-buoyant twisted magnetic flux rope was initially imposed in the convection 
zone just underneath the photosphere, and it started to emerge immediately under 
magnetic buoyancy. The emergence of a subsurface magnetic structure in such a stratified 
atmosphere from the convection zone into the corona gives rise to dramatic expansion 
of the emerging fields. Due to the initial setup of the directions of the emerging 
and ambient fields, the emerging fields in the outer periphery of the flux rope 
reconnect immediately with the ambient open fields, producing a thin column of plasma outflow, 
identified as ``standard jet", following the categorization by \cite{Moore2010}.
Further reconnection also opens up the overlying confining field for the emerging core 
in the flux rope, and promotes the further emergence of the flux rope. In addition, 
strong current builds up within the flux rope and reconnection takes place underneath 
the core field, releasing the twisted core into the corona. The reconnection between 
the core field and the open ambient field drives a violent eruption in the corona, 
generating a much wider column of plasma outflow, known as ``blowout jets", as shown 
by previous simulations \citep{Archontis2013, Moreno-insertis2013}. In particular, 
the outward motion of the ejected plasma in the jet is accompanied with an apparent 
rotating motion \citep{Tian2011}. Comparison of the magnetic twist, upward motion and rotational motion
shows that along magnetic field lines with high twist, the plasma moves upward with a
simultaneous spinning motion \citep{Pariat2009}. Further analysis of the Poynting energy flux 
shows that it is the Poynting flux that drives magnetic energy outward together with mass 
flow driven by Lorentz force and propagation of magnetic twist.

The characteristics of the preceding jet observed in this paper fit most of the theoretical/numerical 
results described above and could be easily identified as a blowout jet. However, the 
identity of the following twin jets and their triggering mechanism seem to be much more 
complicated and unreadable. First of all, no obvious falling-back materials of the preceding 
jet have been detected from the investigation on the {\it STEREO} observations 
(online animation M1 and Figure~\ref{OV}(ds)) at the location of the footpoint region of the right twin 
jet (``Footpoint2" in Figure~\ref{OV}(es)) before its initiation, which indicates that the twins 
should not be triggered by the falling-back materials of the preceding one. Secondly, there was 
not any apparent chromospheric activity detected before the initiation of the twin jets, 
excluding the possibility of simple standard/blowout jets which were triggered by flux emergence or underneath 
shearing/rotational motions. Finally, the twins acted totally simultaneously 
when they were triggered, stayed very close together during their eruption and obtained similar kinetic 
parameters, which inspire us that they should have the same triggering mechanism. 
Due to the lack of underneath magnetic field observations and poor 
spacial/temporal resolutions of the {\it STEREO} EUVI instruments, we are not able to analyze the exact triggering 
mechanism of the twins directly from the observations.

\begin{figure}[tbh!]
\begin{center}
\includegraphics[width=0.9\hsize]{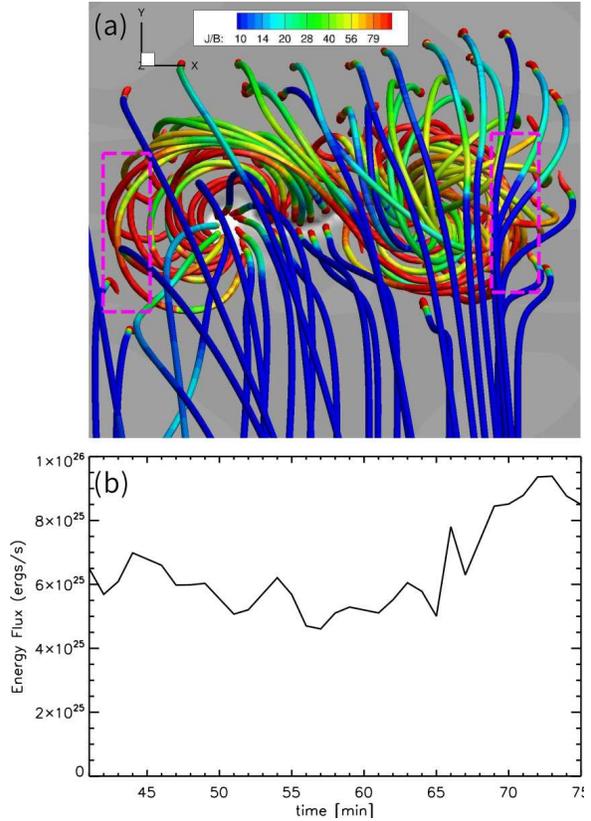}
\caption{(a): 3D structure of the magnetic fields in the simulation around 
the end of the preceding blowout jet at the time of 60 min. 
The grey plane shows the photospheric magnetogram and the rods represents the 
field lines with color showing the $|J|/|B|$ in units of $\mu A\ m^{-2}\ G^{-1}$. (b): the evolution of total energy flux 
at the photosphere.}\label{sigmoid}
\end{center}
\end{figure}

\begin{figure}[tbh!]
\begin{center}
\includegraphics[width=0.9\hsize]{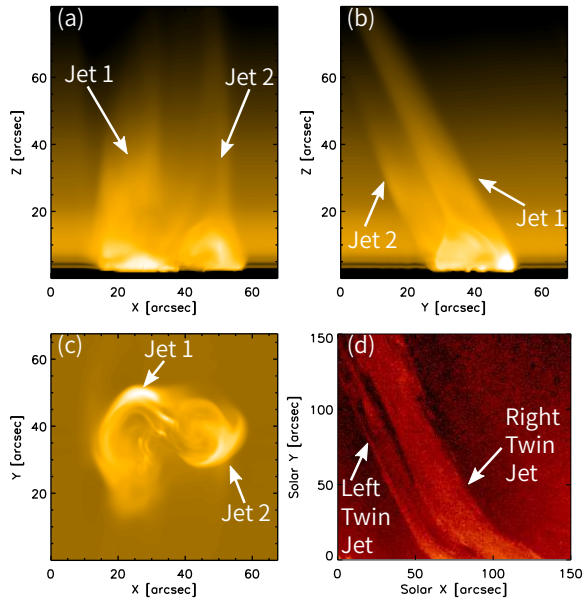}
\caption{Synthetic emission in 171 \AA \  from the simulation at the time of 75 min as viewed from XZ (a), YZ (b) 
and XY (c) plane. An online move (M3) shows the detailed simulated evolution of 
the twin jets as viewed from these three planes. Panel (d) gives the observed twin jets in 
the 304 \AA \  passband of the AIA, the image has been rotated 50\grad{} counter-clockwise.}\label{emission}
\end{center}
\end{figure}

To investigate the possible mechanisms that trigger the twin jets, we continue the simulation in \cite{Fang2014} 
after the eruption of blowout jet. During the blowout jet, sigmoidal current
sheet forms in the emerging flux rope, together with the inverse-S shaped magnetic field configuration, 
shown in Figure~\ref{sigmoid}(a). The reconnected field lines consist of elongated sigmoidal 
magnetic fields embedded in the open fields.
The distribution of the current density $|J|/|B|$ represented by the color of the rods 
clearly shows that the sigmoidal magnetic fields are loaded with strong current in 
the two ends outlined by the two dashed rectangles, as well as the center. 
We also note that the directions of the sigmoidal fields are aligned in the +Y 
direction in the dashed rectangles at the two ends of the sigmoid, which are 
anti-parallel to the ambient fields in the -Y direction. The configuration of the 
magnetic fields and the current in these two regions then 
give rise to reconnection between the sigmoidal and ambient fields at the two ends, 
driving mass outflow along the field lines at the two arms of the sigmoidal structure, 
accompanied with energy flux. As shown by the evolution of total energy flux at the 
photosphere in Figure~\ref{sigmoid}(b), there is a persistent energy flow into the corona from the 
subsurface convection zone during the simulation. To further study the observational 
effect of the reconnection at the two sites, we calculate the synthetic emission from the simulation data and observe the 
structure from multiple points of view, as shown in Figure~\ref{emission}. Panel (a) shows an
X-Z plane view of data, which is perpendicular to the direction of the original flux rope. 
Here two dome structures form in the lower corona, each corresponding to one of the two 
arms of the sigmoidal fields, with one bright column of jet with outflowing plasma extending 
from each dome. Panel (b) is viewed from Y-Z plane, parallel to the axis of the original flux rope, and the two 
simultaneous jets are easily identified here. Panel (c) shows a top view of the domain, 
with a bright sigmoidal shape structure. At the two arms of the sigmoid, there is a significant 
increase of brightness in the emission. The brightening at the two arms results from the 
reconnection of the sigmoidal fields with the ambient fields at the two ends, producing 
two simultaneous jets with energy and mass outflows as shown in Panel (a) and (b). Panel (d) gives the observed twin jets in 
the 304 \AA \  passband of the AIA, which is compared favorably with the simulation in Panel (b). 
A temporal picture of Figure~\ref{emission} showing the entire simulated evolving process could be found in animation M3.

Following the investigation in the simulation results, we propose that the twin jets observed here are probably generated in the same 
fashion, i.e., via the reconnection of ambient fields and a highly twisted sigmoidal magnetic structure, which might result 
from the reconnection during the blowout jets.

\section{Conclusions and Discussions}\label{sect:conc}

In this paper, we presented the first observation and simulation on solar coronal twin jets after a preceding 
solar coronal jet. As to the preceding jet, observations and analysis indicate its blowout jet nature.

Like the jet analyzed in \cite{Liu2014}, the preceding jet underwent an acceleration during its on-disk, 
slow-rise and early off-limb stages, which is widely believed to be majorly introduced by the magnetic reconnections 
between emerging flux and ambient open fields. After then, the jet kept rising under the Lorentz force working. 
Continuous rotational motion of the jet's material around its axis 
could also be observed during the whole off-limb stage. Sine-function fit shows that the rotational motion kept almost the same 
period of about 569$\pm$91 s at different height and did not changed much with time.

Significantly different from the preceding blowout jet, without apparent underneath activities before their initiations, 
the twin jets were triggered around the end of the off-limb stage of the preceding one. They were shot out with pretty 
high axial speeds but rare rotational motions. In lack of the detailed information on the magnetic fields in 
the twin jet region, we instead use a numerical simulation using a 3D MHD model as described in \cite{Fang2014}, 
and find that in the simulation a pair of twin jets form due to reconnection between the ambient open fields and 
a highly twisted sigmoidal magnetic flux which is the outcome of the further evolution of the magnetic fields 
following the preceding blowout jet. Based on the similarity between the synthesized and observed emission we 
propose this mechanism as a possible explanation for the observed twin jets.

Precise estimation and comparison on the energy budgets of the preceding and twin jets would hint us more details of the realistic 
magnetic field configuration and mechanism. However, as we could not determine the mass of the jets and work done by forces 
accurately, this comparison is not able to be done for now. Detailed analysis on the energies from simulation results in the 
future may shed light on this issue. On the other hand, the sigmoidal structure, which was a remarkable 
signature during the twin jets event according to the simulation, was not observed directly in this paper probably due to 
the relatively low cadence and resolution of the {\it STEREO}/EUVI instruments and/or different temperatures. More work in exploring 
and analyzing such twin jets in the future would promisingly lead to improvements.

\acknowledgments{We acknowledge the use of data from AIA instrument on board {\it Solar Dynamics Observatory} ({\it SDO}), EUVI instrument 
on board {\it Solar TErrestrial RElations Observatory} ({\it STEREO}). This work is supported by grants from China 
Postdoctoral Science Foundation, MOST 973 key project (2011CB811403), CAS (Key Research
Program KZZD-EW-01-4), NSFC (41131065 and 41121003), MOEC (20113402110001) and the fundamental research funds for 
the central universities. J.L did part of this work when he was a student visitor at HAO and supported by the 
Chinese Scholarship Council (201306340034). F.F is supported by NASA Grant NNX13AJ04A and the University of Colorado’s George Ellery 
Hale Postdoctoral Fellowship.}

\bibliographystyle{agufull}

\end{document}